\documentclass[twocolumn,showpacs,preprintnumbers,amsmath,amssymb]{revtex4}

\usepackage{graphicx}
\usepackage{dcolumn}
\usepackage{bm}
\usepackage{amsfonts,amssymb,amsmath}

\begin{document}

\preprint{APS/123-QED}

\title{Tunable plasma wave resonant detection of optical beating in high electron mobility transistor}

\date{\today}

\author{J. Torres \footnote {Corresponding Author: torres@cem2.univ-montp2.fr \\ Phone: 33 (0)4 671 449 75}, P. Nouvel, A. Akwoue-Ondo, L. Chusseau}
\affiliation{Centre d'\'Electronique et de Micro-opto\'electronique de Montpellier, UMR 5507 CNRS, Universit\'e Montpellier 2, 34095 Montpellier, France}
\author {F. Teppe}
\affiliation{ Groupe d'\'Etude des Semiconducteurs, UMR 5650 CNRS, Universit\'e Montpellier 2, 34095 Montpellier, France}

\author{A. Shchepetov, S. Bollaert}
\affiliation{Institut d'\'Electronique, de Micro\'electronique et de Nanotechnologie UMR 8520 CNRS, Universit\'e Lille 1, 59652 Villeneuve dÕAscq, France}

\begin{abstract}
We report on tunable terahertz resonant detection of two 1.55 $\mu$m cw-lasers beating by plasma waves in AlGaAs/InGaAs/InP high-electron-mobility transistor. We show that the fundamental plasma resonant frequency and its odd harmonics can be tuned with the applied gate-voltage in the range 75--490 GHz. The observed frequency dependence on gate-bias is found to be in good agreement with the theoretical plasma waves dispersion law.
\end{abstract}

\maketitle

Operating optic-to-electronic data conversion at the terahertz (THz) frequency range is one of the most promising issue of optoelectronic devices. Recently, experimental studies on the plasma resonant detection in high electron mobility transistors (HEMTs) and in a single and a double quantum well field effect transistors (FETs) have been published \cite{knap:2331, knap:675, peralta:1627, teppe:022102, shaner:193507, F.-Teppe:2006uq}. For submicron gate-lengths, fundamental plasma frequencies reach the THz range \cite{Dyakonov:1993sj}. THz detection by plasma waves is easily tunable by changing the gate-voltage. Spectral profiles of THz plasma waves resonances were first reported by T. Otsuji \textit{et al.}, the HEMT being excited by means of interband photoexcitation using the difference-frequency component of a photomixed laser beam \cite{Otsuji:2004ik}. With a similar experiment, we investigate in this letter the plasma waves resonances excited in the channel of HEMTs by the beating of two cw-lasers. We show that the plasma resonant frequencies follow the square-root dependence versus applied gate-voltage as initiallypredicted by Dyakonov-Shur theory \cite{Dyakonov:1993sj}. 

\begin{figure}
\includegraphics[width=0.9\columnwidth]{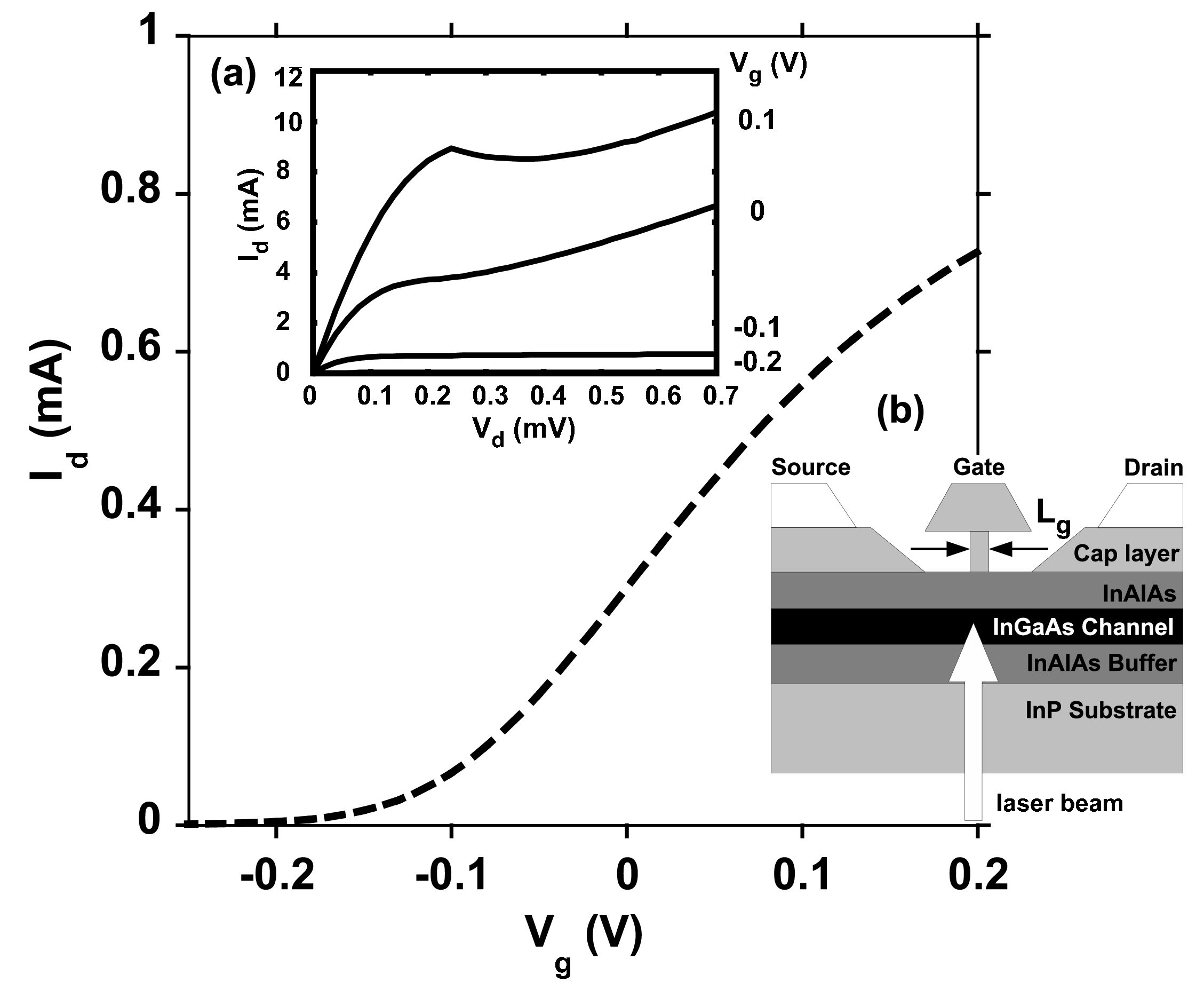}
\caption{\label{fig1} Transfer characteristics of AlGaAs/InGaAs/InP HEMT.  $L_g = 800$\;nm  and $V_{th} = -150$\;mV. Inset (a) : output characteristics at T = 300 K for $V_g = +0.1$\;V, 0 V, -0.1 V and -0.2 V. Inset (b) : schematic of the InGaAs HEMT with the incoming photomixed laser beam by the rear facet.}
\end{figure}

Experiments were performed using an AlGaAs/InGaAs/InP HEMT with gate-length $L_g = 800$\;nm and a threshold voltage of $V_{th}$ = -150\;mV extracted from the transfer characteristics (inset (a) of Fig. \ref{fig1}). The active layers consisted of a 200\;nm In$_{0.52}$Al$_{0.48}$As buffer, a 15 nm In$_{0.7}$Ga$_{0.3}$As channel, a 5-nm-thick undoped In$_{0.52}$Al$_{0.48}$As spacer, a silicon planar doping layer of $6\;10^{12}$\;cm$^{-2}$, a 12-nm-thick In$_{0.52}$Al$_{0.48}$As  barrier layer and a 10-nm-silicon-doped In$_{0.53}$Al$_{0.47}$As cap layer (inset (b) of Fig. \ref{fig1}).

The whole HEMT structure is transparent to the incident radiation excepted the InGaAs-channel where the interband photoexcitation occurs.  By using a tunable optical beating this photoexcitation is modulated over a large frequency range. 
Two commercially-available cw-lasers sources centered at $\lambda_1=1543$\;nm and $\lambda_2=1545$\;nm are used. Each powerful laser ($\approx 20$\;mW) can be tuned over a range of $\approx \pm~1$\;nm by varying the temperature. Their mixing produces a tunable optical beating from 0 to 600\;GHz. The collimated beams are mechanically chopped at 120 Hz and focused onto the HEMT backside using an objective lens (spot size diameter $\approx 5 \; \mu$m). The photoconductivity response, due to the difference frequency generation, is obtained by monitoring the modulation of the dc drain-to-source potential.  \par

\begin{figure}
\includegraphics[width=0.9\columnwidth]{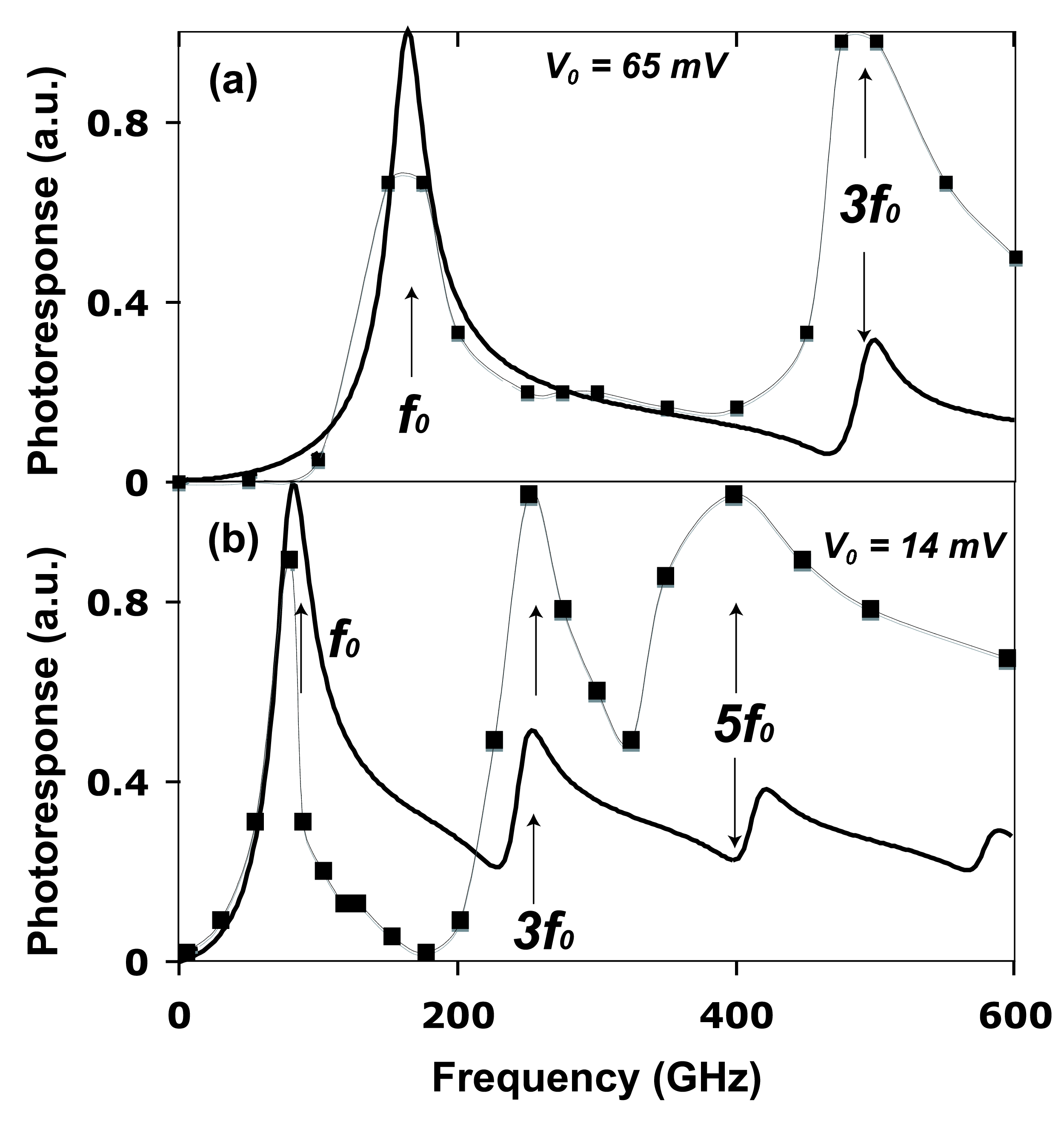}
\caption{\label{fig3} Photoconductive response of sample B vs excitation frequency (0 GHz $\leqslant$ $\Delta f$ $\leqslant$ 600 GHz) for (a) $V_0 = 65$\;mV and (b) $V_0 = 14$\;mV. Squares are experimental data joined by a eye-guidelines. Solid lines are calculation from (\ref{eq:un}). $V_d = 100$\;mV to the source (ground). Arrows is for $f_0$, $3f_0$ and $5f_0$. }
\end{figure}

Figure \ref{fig3} shows the photoconductive response versus excitation frequency at room temperature. The photoresponse intensity normalized to unity is plotted versus the excitation frequency $\Delta f$. Upper and lower panel of (Fig. \ref{fig3}) show the room temperature photoconductive response at $V_0 = 65$\;mV and at $V_0 = 14$\;mV respectively. To discriminate the plasma resonant effect from the background, the photoconductive response under a zero-$\Delta$f-condition was also measured and subtracted from original data. At V$_0$ = 65 mV, two peaks are clearly observed and indicated by arrows. These peaks correspond to the plasma resonance at fundamental frequency ($f_0 \approx 160$\;GHz) and  third harmonic frequency ($3f_0 \approx 490$\;GHz). Like in \cite{Otsuji:2004ik} the peak intensity at 490 GHz is slightly stronger than that at 160 GHz. At V$_0$ = 14 mV, the fundamental ($f_0 \approx 75$\;GHz), third harmonic (3$f_0 \approx 250$\;GHz) and fifth harmonic (5$f_0 \approx 400$\;GHz) plasma frequency peak are observed.  
The drain is biased at $V_d = 100$\;mV. According to the theory devellopped in \cite{veksler:125328}, an accurate description of the resonance response of  HEMTs for $V_d \neq 0$ implies the replacement of the momentum relaxation time $\tau = \mu m^{*}/e$ by an effective momentum relaxation time $\tau_\mathrm{eff}$ given in \cite{Teppe:2005bo} 
\begin{equation}
\label{eq:tau}
\frac{1}{\tau_\mathrm{eff}}=\frac{1}{\tau}-\frac{2v_{0}}{L_g}
\end {equation}
where $v_0$ is the electron drift velocity, $m^{*}$ is the electron effective mass, $\mu = 13000\; \mathrm{cm}^{2}\;\mathrm{V}^{-1}\;\mathrm{s}^{-1}$. With increase of the drain-source voltage, the electron drift velocity increases leading to the increase of $ \tau_\mathrm{eff}$. When  $\omega_0 \tau_\mathrm{eff}$ becomes on the order of unity, the detection becomes resonant; $\omega_0 $ being the fundamental plasma pulsation. The quality factor experimentally obtained from the linewidth [full width at half maximum (FWHM)] of the resonance for $ V_0 = $ 65 mV is $Q = \omega_0  \tau_\mathrm{eff} \approx 1.2$. This value is in good agreement with the calculated values of $ \omega_0 \tau_\mathrm{eff} \approx 1$ obtained 
taking into account the electron drift velocity $v_0 \approx 8~10^{5} \; \mathrm {m.s}^{-1}$ according to Monte Carlo simulation \cite{lusakowski:064307}. In order to determine the theoretical spectral response of plasma waves resonances, the experimental data are compared to the model developed by V. Ryzhii \textit{et al.} \cite{Ryzhii:2002fk} on the plasma mechanism of terahertz photomixing in HEMTs (solid lines in Fig. \ref{fig3}). The responsivity of the HEMT under illumination is
\begin{equation}
R_{\omega}=\frac{e \eta_{\omega}}{\hbar\Omega}\left( \frac{\tan \left( \beta_{\omega}L_g \right) }{\beta_{\omega}L_g}-1 \right)
\label{eq:un}
\end{equation}
where $\hbar\Omega /e = 0.8$\;V in accordance with the incoming radiation wavelengths of $\lambda_1 \approx \lambda_2 \approx 1.55~ \mu$m, $\eta_{\omega}$ determine the excitation of the plasma oscillations by the photogenerated carriers and $\beta_{\omega}$ is the wavevector of the plasma waves given by
\begin{equation}
\label{eq:deux}
\beta_{\omega}^{2}=\frac{\omega(\omega+i\nu)}{s^{2}} \qquad s=\sqrt{\frac{eV_0}{m^*}}
\end{equation}
$s$ being the plasma waves velocity and $\nu$ the collision frequency of carriers in the absorption layer ($\nu~= 1.5 \; 10^{11} \; \mathrm{s}^{-1}$). As given in \cite{Ryzhii:2002fk}, $\eta_{\omega}$ is a function of the carrier relaxation times $\tau_\mathrm{eff} $ and of the beating pulsation $\omega$. 
As a result, the FWHM of the plasma waves resonances obtained by the calculation with $\omega_0\tau_\mathrm{eff}  = 1$ agree well with experimental data which are spectrally broaden because of the lower quality factor of the THz  ``cavity''. The calculated resonant frequencies of the plasma waves obtained from this model are seen in good accordance with experiments in Fig. \ref{fig3}. It should be emphasized that these calculated results are obtained without any adjustable parameter. Only InGaAs material parameters such as electron effective mass ($m^{*} = 0.041~m_0 $) and HEMT parameters ($L_g, \tau_\mathrm{eff} $) have been introduced in the calculation. A small discrepancy between model and experiments deals with the photoresponse peak amplitudes. They are nearly constant in experiments, whereas they are expected to decrease in the model when $\Delta f$ increases. A possible explanation is that the model calculate the HEMT responsivity from the ac gate-channel current  while we experimentally monitored the dc drain-source voltage. The transfer function between both being the $I_d$--$V_d$ characteristic of the HEMT that is spectrally dependent.

\begin{figure}
\includegraphics[width=0.9\columnwidth]{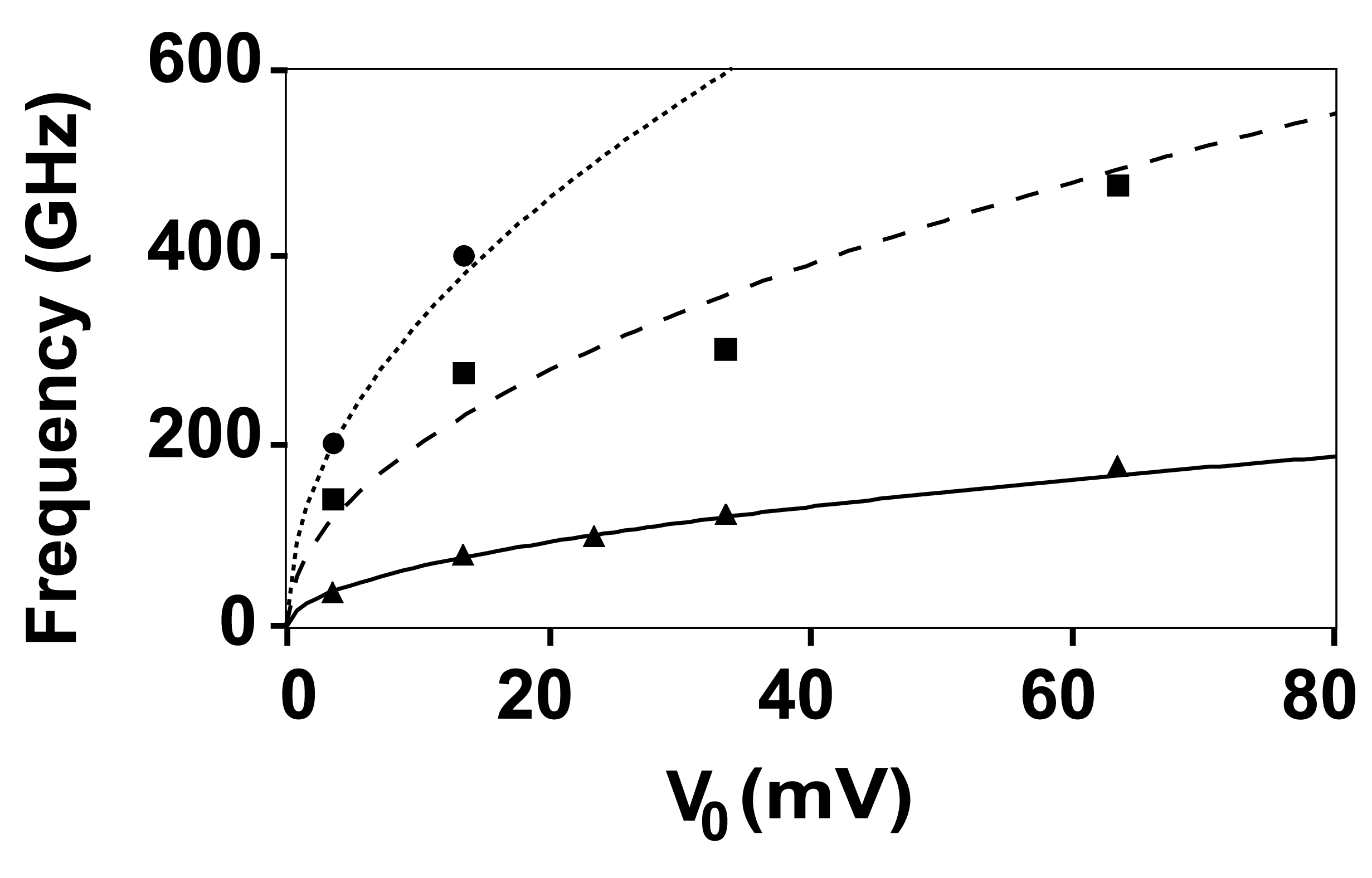}
\caption{\label{fig4} Frequency dependence of the maxima of the plasma resonant peak versus swing voltage. Experiments : $f_0$ ($\blacktriangle$), 3$f_0$ ($\blacksquare$) and 5$f_0$ ($\bullet$). Lines are calculations using (\ref{eq:trois}) with $L_g = 800$\;nm. Solid line denotes for $f_0$, dashed line for 3$f_0$ and dotted line for 5$f_0$.}
\end{figure}

The frequency dependence of the plasma waves is plotted in Fig. \ref{fig4} versus the swing-voltage. The experimental data for the fundamental frequency and its odd harmonics are obtained by picking-up the frequency position of the plasma-wave resonance maxima in Fig. \ref{fig3} for several values of $V_g$. Lines are calculations using  \cite{Dyakonov:1993sj} 
\begin{equation}
f_n = \frac{2n+1}{4L_g} s 
\label{eq:trois}
\end{equation} 
\\
where n = 0, 1, 2, \dots, and $L_g = 800$\;nm. Theoretical calculations well support experimental datas for fundamental plasma waves frequency and its odd harmonics. 

In conclusion, the excitation of plasma waves in AlGaAs/InGaAs/InP HEMT channel was performed at room temperature by using a pair of commercially available cw-lasers delivering a THz optical beating.
We show that the fundamental plasma resonant frequency and its odd harmonics can be tuned with the applied gate-voltage in the range 75--490 GHz and follow the predicted square-root behaviour. A good quantitative agreement between experiments and Dyakonov-Shur theory is obtained. 

Future works will involved shorter gate-length devices to reach higher plasma frequencies thus allowing ultrahigh optoelectronic data conversion. 

\textbf{ACKNOWLEGMENTS}

This work was partly supported by CNRS grant and special funding.

\end{document}